# Generation of a retro-reflected wave by interaction of an evanescent wave with a sub-wavelength structure

Young-Gu Ju,[1,*] Thomas Milster[2]

[1]*Department of Physics Education, Kyungpook National University, Daegu 702-701, Korea*
[2]*College of Optical Sciences, The University of Arizona, 1630 East University Boulevard, Tucson, Arizona 85721, USA*
[*]*ygju@knu.ac.kr*

**Abstract:** Numerical calculations were performed to examine the mechanisms for generation of a retro-reflected wave from the interaction of an evanescent wave with a sub-wavelength structure using the finite-difference time-domain (FDTD) method. The simulation shows that an evanescent wave is reflected from the structure at the interface between a high index dielectric material and a low index material. The reflected evanescent wave couples into the upper medium and radiates its energy forming a retro-reflected wave, which appears as a sharp peak near the edge of the structure when imaging the structure in hyper-numerical-aperture solid immersion microscopy. We propose a simple theory and verify it through FDTD calculation under various circumstances in order to explain peculiar features of this phenomenon. Furthermore, we suggest a way to control the reflection of the evanescent wave by taking advantage of the interference of the evanescent wave inside the structure.

## 1. Introduction

Recently, nano-science and nano-technology are of high interest because of their significant impact on various fields such as semi-conductor technology [1], material science [2], photo voltaics [3], chemistry [4] and biology [5]. Advances in nano-technology also demand a strong need for measurement technology in order to investigate nano-size features. Traditional optical microscopy has a fundamental limitation in resolution, due to diffraction, which specifies that the minimum resolution is proportional to wavelength and inversely proportional to numerical aperture. One way of improving resolution is to increase the refractive index of the medium, since the numerical aperture is the product of the refractive index of the medium and the sine of the aperture angle. A solid immersion lens(SIL) is an example utilizing a high index medium, such as silicon, which has an index of approximately 3.5 in the infrared spectrum [6]. As is well known in the semiconductor industry, the use of oblique illumination can improve resolution, since oblique incidence shifts the spatial frequency range and can allow about two-fold enhancement in the resolution limit. However, use of the oblique incidence technique also causes fringes to appear at the illuminating side of the pattern edges in a subwavelength feature [7, 8]. In this paper, we analyze this problem by calculating the electro-magnetic fields generated in a SIL microscopy geometry. The simulation is carried out by means of FDTD(Finite-difference-time-domain), which is known to be accurate and efficient for analyzing nano-scale structure [9, 10]. Results indicate that image features are related to the evanescent wave propagating in the gap under the SIL and its interaction with the pattern under observation, where a retro-reflected plane wave is generated. In addition, a simple theory is suggested to explain this phenomenon and demonstrate how to control this evanescent wave reflection by changing the structural parameters of the pattern.

## 2. Simulations

FDTD calculations are performed in order to simulate imaging of a hyper-numerical aperture SIL microscope. The software used for simulation is Sim3D_MAX [11]. In the simulation, the SIL is assumed to have an index of 3.5, which is close to that of Si(silicon). The refractive index distribution in the xz plane used for calculation is displayed in Fig. 1. The upper region represents the incident Si medium. The two rectangles at the bottom of the Si correspond to two subwavelength metallic lines placed on the Si substrate. This configuration assumes a SIL in close contact with a Si substrate that has metallic lines on the opposite side, which is similar to the geometry described in reference [8]. Since the SIL and substrate are the same material, they can be considered as a single medium. A Debye model inside SIM3D was used to calculate the permittivity of the metal. The refractive index profile in Fig. 1 shows that metal has index of -3, which is an arbitrary index to represent a special material described by a separate dispersion function. In this case, the metal is aluminum. As for the rest part of refractive index profile, the lower medium comprises vacuum with refractive index of 1.0.

For simplicity of analysis, the configuration has symmetry in the y direction, which implies that the two rectangles represent two long metallic lines in the y-direction. In addition to the index profile, the FDTD simulation requires constraints on the light source. Since hyper-numerical aperture SIL microscopy is simulated, an obliquely incident plane wave is used. Perfectly matched layers(PMLs) are used for all boundaries. The plane wave source is placed as close to the interface as possible to maintain the uniformity of the incident wave in the middle region by avoiding edge effects. Although the obliquely incident wave favors the Floquet boundary condition on the sides for obtaining better uniformity, the non-periodic geometry of the object conflicts with this boundary condition. Use of the PML boundary condition gives some non-uniformity in the incident wave, especially near the edge of the calculation space, but it does not affect the analysis significantly, since the objects and the electromagnetic fields of concern are located in the central part of the geometry. Lastly, the wavelength of source used for calculation is 1.0 μm. Therefore, effective wavelength inside the Si material is 286 nm. Since the absorption coefficient of Si is less than 100 cm$^{-1}$ at this wavelength, the light can transmit more than 100 µm without substantial loss [12]. If longer wavelength is necessary for higher transmission, the feature size should increase with the same factor in order to maintain the same result according to the scalability of Maxwell's equations.

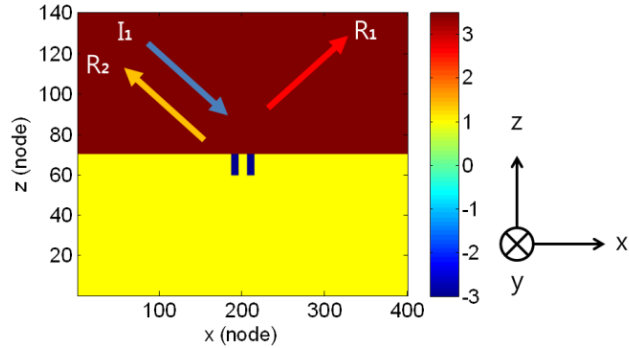

Fig. 1. The refractive index profile used for FDTD calculation shows the two metallic lines under Si substrate exposed to the air. The indices of Si and air are 3.5 and 1.0, respectively. A node is 10 nm. I1, R1 and R2 represent the incident wave, specular reflection and retro-reflection, respectively.

Dimensions of the subwavelength metallic lines are 90 nm in width, 100 nm in height, and 90 nm for the spacing between them. The angle of incidence in the Si is 50 degrees, and light is polarized in the y direction, which is TE(transverse electric) polarization. TE polarization is chosen for the simplicity, since TM polarization exhibits more complex behavior. However, artifacts described here occur for both states of incident polarization. The numbers of nodes used for calculations are 400 X 100 X 140 in x, y and z directions, respectively. One node corresponds to 10 nm.

## 3. Results and Discussion

Monitoring the electric fields for a certain amount of time in FDTD provides the amplitude and the phase information at each location. Amplitude of the electric field near the object and at the image are calculated, and the results are presented in Fig. 2. Fourier transforming the near field and filtering the angular spectrum within a given numerical aperture is followed by an inverse Fourier transform in order to give the field at the image. The numerical aperture(NA) used in the calculation corresponds to a 60 degree cone, or NA=3.03. In spite of the finite numerical aperture, the image doesn't lose detail of the near field. Two valleys observed in Fig. 2-(a) represent the two metallic lines. The two valleys are well separated and discernible, implying that the resolution is smaller than 90 nm($\lambda/11$). More simulation reveals that the limiting resolution is about 70 nm($\lambda/14$), which is slightly better than the theoretical limit given by diffraction formula under oblique illumination($\lambda/(4\,n\,\sin\theta)$). The peculiar feature observed in Fig. 2 is the existence of large and oscillatory fringe near the edge of the first metallic line. Since it is not the part of the object feature, it is an artifact of the image coming from an unknown origin. These fringes are also observed experimentally in a previous publication [8]. Therefore, these peaks are not an erroneous outcome of the simulation, but they are due to the nature of imaging in this geometry.

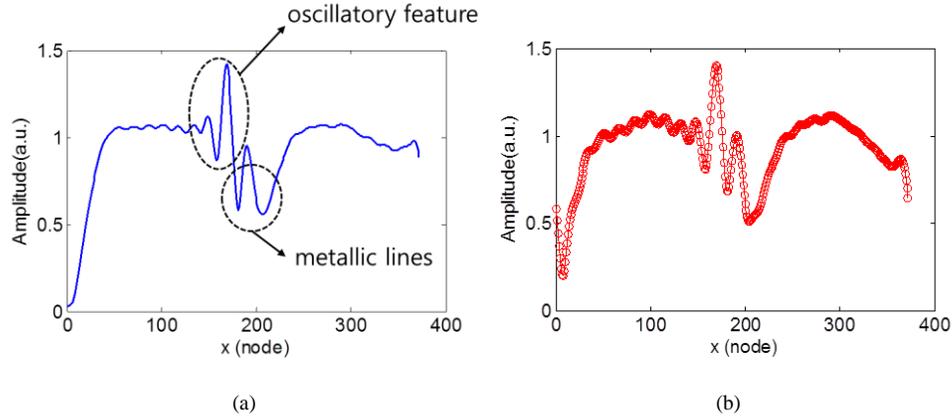

Fig. 2. The electric field profile for two metallic lines taken (a) near object and (b) at the image plane of an objective lens.

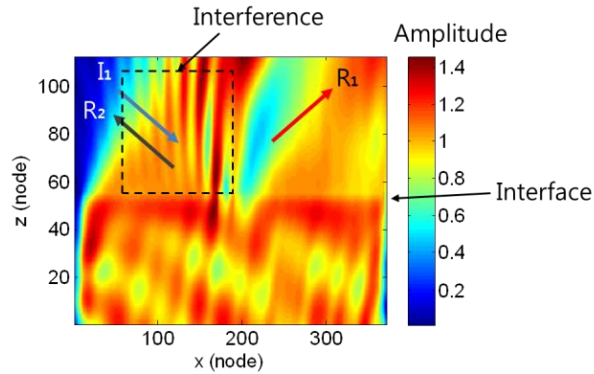

Fig. 3. Subtracting the incoming wave from the calculated electric field provides the cross-section of the amplitude of the outgoing electric field in xz plane.

For further analysis, a cross-section of the electric field is captured and processed to remove the incoming wave. Post-processing is carried out by subtracting the incoming reference wave from the field distribution. Calculation of the reference wave resulted from monitoring the incident wave propagating in uniform Si. The post processing illustrates properly the retro-reflected wave from the object and the interfaces, although it could give false data below the interface. Figure 3 displays the amplitude distribution of the processed data. On the left side of the object in the upper medium, there are straight lines indicating interference between the retro-reflected wave and the specular reflection. Since the interference fringes are vertical, the two waves have opposite propagation vectors in the x direction. The retro-reflected wave originates from the interaction between the evanescent wave in the vacuum and the subwavelength feature.

Two unusual features are significant in this phenomenon. Firstly, the reflection occurs despite the absence of an object in the upper Si region. Secondly, the retro-reflected wave has opposite signs of the propagation vectors both in x and z directions compared to the incident wave. The second fact indicates that the reflected wave has the nature of a retro-reflection, instead of exhibiting scattering over a large angle. Since the feature size of 90 nm is much smaller than the wavelength in the medium(286 nm), scattering with wide angular distribution might be expected, rather than a retro-reflection.

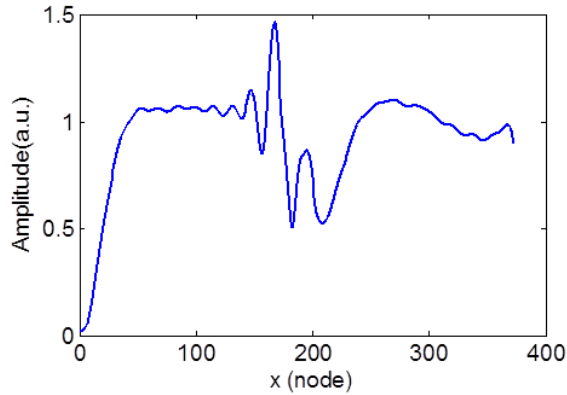

Fig. 4. The use of artificially high index for the line pattern shows similar results to that of the metallic lines. The refractive index of the dielectric lines used in the simulation is 30.

In order to check if this phenomenon is related to plasmons, the metallic lines are replaced with dielectric lines. To make the situation similar to the metallic case, the refractive index of dielectric lines is set to 30. The use of a huge refractive index is artificial, but effective, in making a highly reflective surface without causing a plasmonic effect. The result from the dielectric lines of high index is shown in Fig. 4 and resembles that of the metallic lines shown in Fig. 2-(a). The similarity between the two results implies that the phenomenon is not related to plasmons. Besides, plasmonic effects usually happen for TM(tansverse magnetic) polarization, which is not the case of this simulation.

Organizing the results from the simulations leads to a conjecture that the evanescent wave plays an important role in producing retro-reflection from an object. The evanescent wave formed beneath the interface doesn't propagate in the z direction perpendicular to the interface. However, it can propagate in parallel to the interface in x and interact with the object at the bottom of the surface, giving rise to an evanescent reflection. The reflected evanescent wave forms a standing wave with the incident evanescent wave and transfers its phase variation to a propagation vector in the upper Si medium through boundary conditions. At the interface, the y-polarized electric field in the lower region coincides with that of the upper region. In this manner, the reflected evanescent wave induces a retro-reflected wave in the upper medium. The generated retro-reflected wave has a propagation vector with reversed signs both in x and z directions with respect to the incident wave. As a result, the reflected evanescent wave radiates its energy into the upper medium through strong coupling to the retro-reflected wave. The loss of energy appears as an attenuation of the retro-reflected wave. Therefore, the total near-field reflection forms a localized standing wave near the left edge of the object due to interference with the specular reflection. This conjecture explains the strong satellite peaks and the oscillatory behavior near the illuminating side of the pattern.

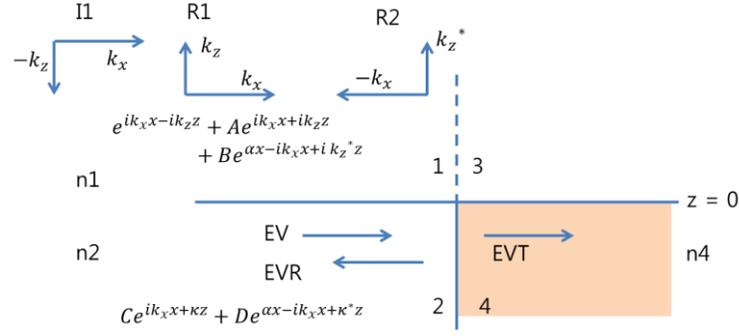

Fig. 5. The schematic diagram illustrates the generation of conjugate wave by means of the reflection of the evanescent wave. I1, R1 and R2 represent an incident wave, the reflected wave and the retro-reflected wave, respectively. EV, EVR and EVT also represent the evanescent wave, the reflected evanescent wave and transmitted evanescent wave, respectively.

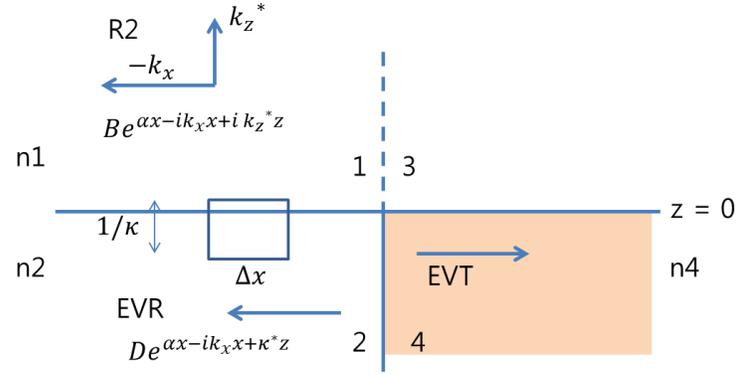

Fig. 6. Poynting theorem provides a way to estimate the decay constant of the evanescent standing wave near the structure.

The schematic diagram of Fig. 5 illustrates the concept of evanescent retro-reflection. The medium in the lower right corner replaces the subwavelength object from the simulation and simplifies the geometry for theoretical work. The reflected wave comprises a range of spatial frequencies both in x and z directions, since the reflected field is localized near the left-hand side of the object. The z-dependent amplitude B indicates that the reflected wave is not a plane wave, but a localized wave packet consisting of a range of spatial frequencies. However, the interference fringe, as seen in Fig. 3, shows a periodicity to some extent, which implies that the reflected wave has a narrow spatial frequency bandwidth. Figure 3 also shows that the reflected wave well above the interface maintains the same periodicity as observed near the interface, which implies that $k_z^*$ is close to $k_z$. Therefore, the reflected wave is a retro-reflected wave having a k-vector with opposite directional signs to that of the incident wave.

As for the decay constant $\alpha$ of the reflected wave, it should satisfy the Poynting theorem of energy conservation, $\nabla \cdot \vec{S} = 0,$ in a small region. Therefore, the net energy flux through the closed box drawn in Fig. 6 should be zero. Since the evanescent wave doesn't propagate in the negative z direction, the fluxes through the sides and the top boundary are included in the calculation. Equation (1) describes how $S_x$ and $S_z$, the x and z components of the Poynting vector, respectively, are related. The height of the box is about $1/\kappa$, which is the approximate depth of the evanescent wave. Substituting $1/\kappa$ for $\Delta z$ gives Eq. (2).

$$\frac{\partial S_x}{\partial x}\Delta z = -S_z \qquad (1)$$

$$\frac{\partial S_x}{\partial x} = -\kappa S_z \qquad (2)$$

$S_x$ derives from the electric field of the evanescent wave while $S_z$ derives from the electric field above the interface. $S_x$ and $S_z$ are now expressed in terms of the electric fields. $E_{y1}$ and $E_{y2}$ correspond to fields in the y direction taken from the medium 1 and medium 2, respectively.

$$S_x = \frac{1}{2}Re\left(\frac{E_y\frac{\partial E_y^*}{\partial x}}{i\omega\mu_0}\right) = -\frac{1}{2}k_x\frac{|E_{y2}|^2}{(\omega\mu_0)} \qquad (3)$$

$$S_z = \frac{1}{2}Re\left(\frac{E_y\frac{\partial E_y^*}{\partial z}}{i\omega\mu_0}\right) = \frac{1}{2}k_z\frac{|E_{y1}|^2}{(\omega\mu_0)} \qquad (4)$$

Since $E_y$ is assumed to exponentially decay as it goes in the negative x direction, the square of $E_y$ follows the relation shown in Eq. (5). Moreover, the boundary condition states that the parallel components of the electric field are continuous across the boundary, which allows the relation of Eq. (6).

$$|E_y|^2 \propto e^{2\alpha x} \qquad (5)$$

$$|E_{y1}|^2 = |E_{y2}|^2 \qquad \text{at z=0} \qquad (6)$$

Substitution of Eq. (3) and Eq. (4) into Eq. (2) and application of the Eq. (5) and Eq. (6) results in the simple relation

$$\alpha \simeq \frac{\kappa k_z}{2k_x} = \frac{\kappa}{2\tan\theta} \sim \frac{\kappa}{2}, \qquad (7)$$

where $k_z^*$ is similar to $k_z$. Equation (7) states that the decay constant $\alpha$ can be described by the decay constant of the evanescent wave and the propagations vectors of the incident wave. Although Eq. (7) shows dependence on the incidence angle $\theta$, it can be misleading because decay constant $\kappa$ also has an angle dependence. With an incident angle of 50 degrees, $\alpha$ is roughly $\kappa/2$.

Periodicity and decay constant of fringes observed near the object are analyzed in order to see if they relate to the standing wave coming from reflection of the evanescent wave. The phase distribution of the incident wave measured in the x direction is used to define periodicity of the incident wave, as shown in Fig. 7-(a). The measured periodicity is 38 nodes, which corresponds to 380 nm. Half of the periodicity in x direction is 190 nm, which is the expected periodicity of the evanescent standing wave. According to Fig. 7-(b), distance between the two neighboring peaks in the fringes is 19 nodes, or 190 nm. This agreement supports the conjecture that the observed fringes are due to the standing wave formed from the incident and reflected evanescent waves.

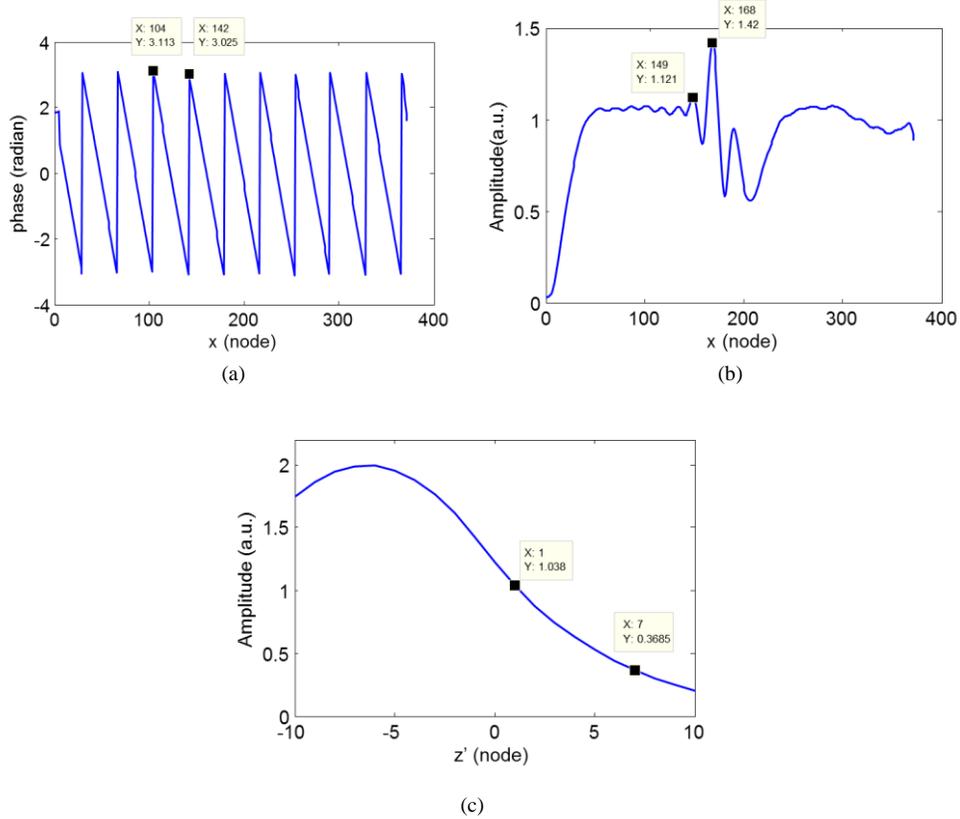

Fig. 7. The period and the decay constant of the fringes near object are extracted and compared with the theoretical values:(a) the phase plot of the reference wave, (b) the coordinates of the two peaks of the fringes near the structure and (c) the field amplitude across the boundary in z direction. z'=56-z and x = 50 nodes.

Calculation of the decay constant $\alpha$ requires measurement of decay constant $\kappa$ from the evanescent wave according to Eq. (7). Cross-section of the z-direction electric field across the interface is presented in Fig. 7-(c). Since z=0 corresponds to the interface, the two points after this point allows calculation of the decay constant $\kappa$, which is Log(1.038/0.3685)/(7-1) or 0.16 (1/node). In the same way, the coordinates of two peaks in the fringes, as shown in Fig. 7-(b), provides the decay constant $\alpha$. Since the center of the oscillation is 1.05, this value is subtracted from the vertical coordinate to calculate the correct amplitude. The measured $\alpha$ is Log((419-50)/(121-50))/19 or 0.087 (1/nodes). Since the measured $\alpha$ is about half of $\kappa$, it agrees reasonably well with Eq. (7). This consistency also reinforces the idea of a reflected evanescent wave.

The change of the two metallic lines into a single metallic line doesn't affect the shape of fringes near the object. Figures 8 and 9 show the evanescent field profiles and the near-field profiles, respectively. The evanescent field is taken right below the interface, while the near-field was taken right above the interface. Although the evanescent wave displays the same peak near the object as the near field, the field amplitude is small and suffers from more noise of the calculation compared to that of the near field. The differences between two metallic lines and a single line are ignorable. It suggests that the evanescent wave mainly reflects from the first surface of the object it meets. Therefore, the fringe is not an artifact or an interference coming from the double metallic patterns.

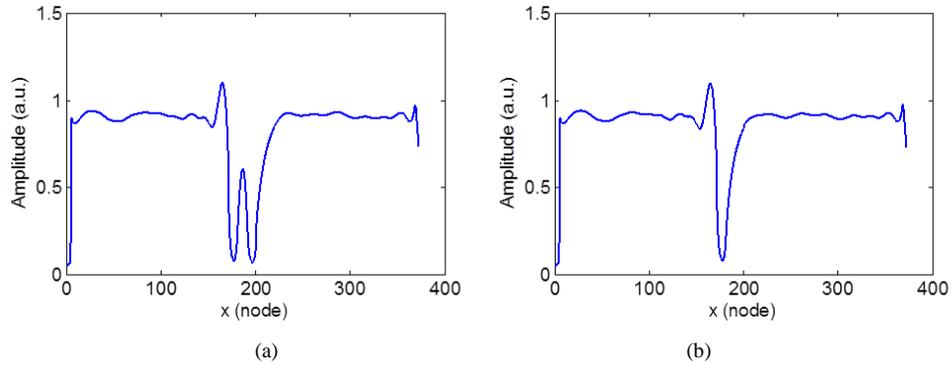

Fig. 8. The evanescent field profile from (a) two metallic lines and (b) single metallic lines.

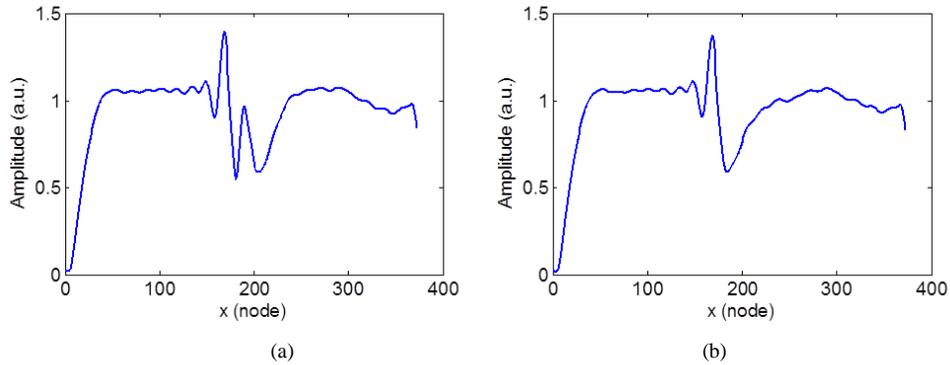

Fig. 9. Near field profile from (a) two metallic lines and (b) single metallic lines.

The theory of the evanescent reflection can be used to engineer reflective properties. Figure 10 illustrates how to control evanescent reflection by adjusting the width of a dielectric line. The basic concept is to vary interference between reflections from the first surface and the second surface of the dielectric object. Since the evanescent wave is similar to a guided wave, adjusting the width can produce constructive interference or destructive interference from the two reflections. A large difference from classical interference is the insensitivity on the dielectric material refractive index. Since $k_x$ of the evanescent wave is determined by $k_x$ of the incident wave, the interference of the evanescent wave doesn't depend on the refractive index of the object, as long as the object itself doesn't significantly perturb the evanescent wave.

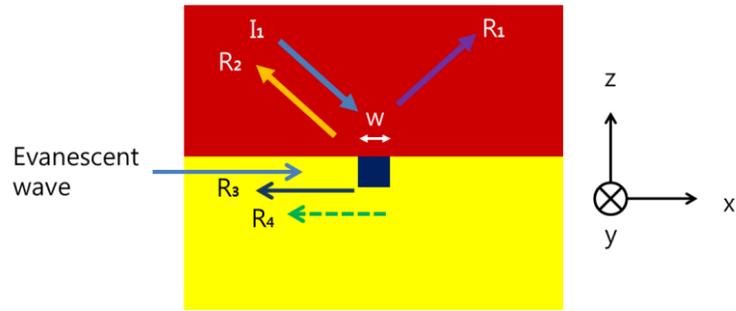

Fig. 10. Schematic diagram illustrates the control of the reflection of the evanescent wave by modifying the dielectric structure such as the width w. R3 and R4 represent the evanescent reflection at the first surface and the second surface, respectively.

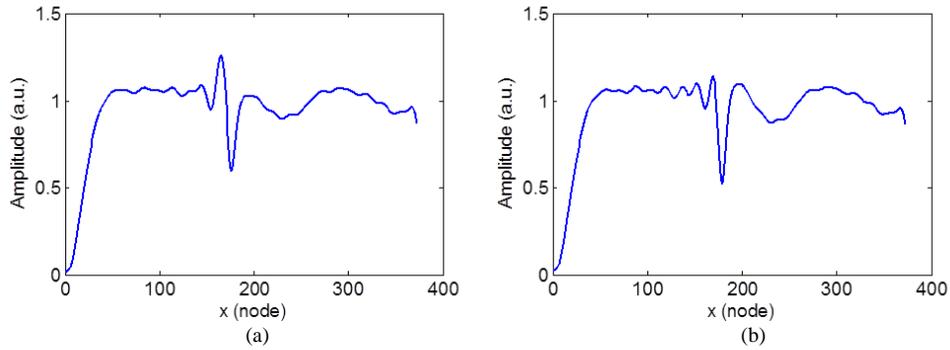

Fig. 11. The near field profile calculated from (a) HR condition( w = 100 nm) and (b) AR condition( w = 190 nm). The index of the dielectric is 3.0.

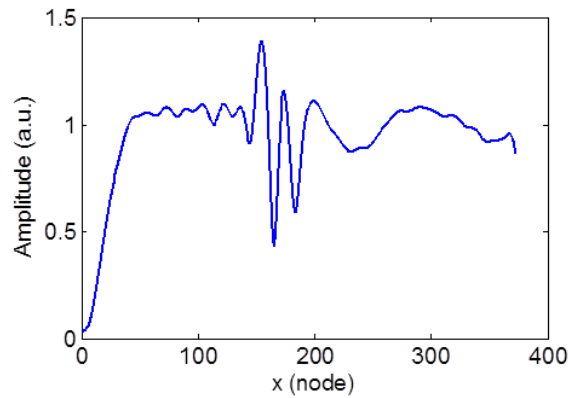

Fig. 12. The near fields profile from two periods of dielectric lines is displayed. The index of the dielectric is 3.0. Both the width and the spacing between the lines is 100 nm.

In order to demonstrate the idea of evanescent wave interference, width of a single dielectric line is varied, and results are presented in Fig. 11. The dielectric refractive index used in the simulation is 3.0. Since the phase periodicity of the incident wave is 38 nodes or 380 nm, 100 nm width corresponds to a quarter-wave, which induces constructive interference due to the opposite phase change at the two interfaces. On the other hand, 190 nm corresponds to a half wave condition, leading to a destructive interference. The peak amplitude for 190 nm is much smaller than that for 100 nm, which agrees with the concept of evanescent interference. Comparing the case of 190 nm with that of the single metallic line, as seen Fig. 9-(b), the reduction of the fringe due to the interference is obvious, while the peak at 100 nm width is similar to that of the metallic line.

A more ambitious attempt to enhance the evanescent reflection is attempted by increasing the number of lines. At first one line is added, for a total of two lines. The width and the spacing between the two dielectric lines are 100 nm, since the $k_x$ is about the same over this region. The result, as seen in Fig. 12, shows that peak amplitude increased from 1.25 to 1.4. However, adding more lines does not increase peak amplitude. This result is ascribed to loss of the guiding effect over a longer distance.

## 4. Conclusions

FDTD analysis of hyper-numerical aperture imaging reveals existence of strong and oscillatory fringes near the object under oblique illumination. The fringe patterns are nearly straight and normal to the interface. The replacement of metallic objects with high index dielectric objects makes negligible difference in the fringe pattern, which indicates that this phenomenon does not relate to a plasmon effect. The directionality of the fringes also indicates that the reflected wave is a retro-reflected wave having propagation vector components with opposite signs to those of the incident wave. Formation of the fringes is attributed to a standing wave between the incident wave and the retro-reflected wave. Although the object is located at the bottom of high index medium, it causes the reflection by means of interaction with an evanescent wave. The evanescent wave travels parallel to the interface and reflects from the side of the object. The reflected evanescent wave also transfers its phase modulation into the upper medium and generates a retro-reflected propagating wave. The evanescent wave loses its energy through coupling to the propagating wave. In this manner, the evanescent wave suffers from attenuation, which is also observed in fringes near the object. We derive a simple formula for the decay constant of the reflected evanescent wave, which relates to the decay constant of the evanescent wave in the vertical direction. Measurement of phase variation of the incident wave verifies that the fringe pattern near the object is a standing wave between the incident wave and the retro-reflected wave. In addition, the decay constant from the derived formula agrees with the measured value from the fringe pattern, which also supports the proposed concept. Furthermore, we suggest a way to control the reflection of the evanescent wave by taking advantage of interference between the evanescent waves inside a structure. Simulations demonstrate enhancement or suppression of the evanescent reflection by adjusting the width of the dielectric object or the number of objects. In this way, the concept of evanescent wave retro-reflection can be applied to explain many imaging behaviors observed under hyper-numerical aperture microscopy using a SIL.

**Acknowledgement**
This research was supported by Basic Science Research Program through the National Research Foundation of Korea(NRF) funded by the Ministry of Education, Science and Technology(2011-0007648). This research was supported by Kyungpook National University Research Fund, 2012. This research was supported by the fund from Kyungpook National University Specialization Program, 2012.